\documentclass{iopart}

 \def\b{\beta }  
\def\d{\delta }

 \def\t{\theta }

\def\beq{\begin{equation}}   \def\eeq{\end{equation}}
\def\la{\langle}  \def\ra{\rangle}

\def\bea{\begin{array}}   \def\ea{\end{array}}
\def\beqn{\begin{eqnarray}}   \def\eeqn{\end{eqnarray}}
\def\dg{\dagger}

\font\got=eufm10 scaled \magstep1
\font\gotscr=eufm7 scaled \magstep1
\font\gotscrscr=eufm5 scaled \magstep1
\newfam\gotfam
\textfont\gotfam=\got
\scriptfont\gotfam=\gotscr
\scriptscriptfont\gotfam=\gotscrscr
\def\got{\fam\gotfam}

\font\Bbb=msbm10 scaled \magstep1
\font\Bbbscr=msbm7 scaled \magstep1
\font\Bbbscrscr=msbm5 scaled \magstep1
\newfam\Bbbfam
\textfont\Bbbfam=\Bbb
\scriptfont\Bbbfam=\Bbbscr
\scriptscriptfont\Bbbfam=\Bbbscrscr
\def\Bbb{\Bbbfam}

\font\Cal=msbm10 scaled \magstep1
\font\Calscr=msbm7 scaled \magstep1
\font\Calscrscr=msbm5 scaled \magstep1
\newfam\Calfam
\textfont\Calfam=\Cal
\scriptfont\Calfam=\Calscr
\scriptscriptfont\Calfam=\Calscrscr
\def\Cal{\fam\Calfam}

\def\gappeq{\mathrel{\rlap {\raise.5ex\hbox{$>$}}
{\lower.5ex\hbox{$\sim$}}}}

\def\lappeq{\mathrel{\rlap{\raise.5ex\hbox{$<$}}
{\lower.5ex\hbox{$\sim$}}}}

\begin{document}

\title{ Multi-particle correlations in $qp$-Bose gas model }

\author{ L V Adamska and  A M Gavrilik  }
\address{Bogolyubov Institute for Theoretical Physics,
         03143 Kiev, Ukraine }
  \ead{omgavr@bitp.kiev.ua }        

\begin{abstract}
The approach based on multimode system of $q$-deformed oscillators
and the related picture of ideal gas of $q$-bosons enables to
effectively describe the observed non-Bose type behaviour,
in experiments on heavy-ion collisions, of the intercept
(or the "strength") $\lambda$ of the two-particle
correlation function of identical pions or kaons.
In this paper we extend main results of that approach in 
the two aspects: first, we derive in explicit form the 
intercepts of $n$-particle correlation functions in the case 
of $q$-Bose gas model and, second, provide their explicit
two-parameter (or $qp$-) generalization.
\end{abstract}



\section{Introduction}


Quantum and $q$-deformed algebras are known to be 
very useful in diverse problems in many branches 
of mathematical physics and modern field theory        \cite{Bied,Zachos},
as well as in molecular/nuclear spectroscopy           \cite{Chang}.
As well fruitful should be their direct 
application in the phenomenology of particle 
properties  (see                                    \cite{Chai,Fair,JPA,GI},
and also                                              \cite{NP_B}
with references therein).
Recently, it has been demonstrated                         \cite{AGI-1}
that the use of multimode $q$-deformed oscillator
algebras along with the related  picture of ideal gas
of $q$-bosons ($q$-Bose gas model) proves its efficiency
in modelling the unusual properties of the intercept
$\lambda$ of the two-particle correlation function, that is 
the measured value corresponding to zero relative momentum 
of two identical mesons, pions or kaons, produced and
registered in relativistic heavy-ion collisions              \cite{Heinz99},
where $\lambda$ exhibits sizable observed 
deviation from the naively expected purely 
Bose-Einstein type behaviour.
The model predicts                                        \cite{AGI-2,AGP},
for a fixed value of $q$,
the exact shape of dependence of the intercept
$\lambda = \lambda({\bf K})$ on the pair mean momentum
${\bf K}$ and suggests asymptotic coincidence of
$\lambda_\pi$ and $\lambda_K$ of pions and kaons.
Put in another words, the intercept $\lambda$,
being connected directly and unambiguously with the
deformation parameter $q$, tends in the limit of large 
pair mean momentum to a constant, lesser than unity,
determined just by $q$ and shared by pions and kaons.
It is worth noting that confronting the predicted 
$\lambda_\pi$ behavior with the corresponding data 
from STAR/RHIC shows nice agreement                           \cite{AGP}, 
at least in the case of two-pion correlations.

While two-particle correlations are known to carry
information about the space-time structure and 
dynamics of the emitting source                             \cite{Heinz99}, 
in connection with some recent experiments
it was pointed out                                     \cite{Adams,Heinz97}
that taking into consideration, in addition
to the single particle spectra and two-particle 
correlations based analysis, the amount of data 
concerning 3-particle correlations provides an 
important supplementary information on the properties 
of the emitting region, valuable for confronting 
theoretical models with concrete experimental data. 
Likewise, study of $4$- and $5$-particle correlations 
is also desirable                                            \cite{Csorgo}.
All that motivates the main goal of present 
contribution that is to derive the explicit formulas 
for the intercepts of higher order ($n$-particle, with $n\ge 3$) 
HBT correlations.
Moreover, below we will obtain in explicit form 
the intercepts $\lambda^{(n)}$ of $n$-particle 
correlations for the extended version of the developed 
approach when one uses the two-parameter $qp$-deformation 
of bosonic oscillators and, respectively, the model 
of gas of $qp$-bosons.

From the very first, and up to more recent, 
applications of the $q$-algebras to phenomenology of 
hadrons there is growing evidence                           \cite{JPA,NP_B}
that the phase-form of $q$-parameter is
of great importance.
Therefore, we hope that possessing the most general
formulas for $n$-particle correlations and confronting
them with the data from contemporary experiments will be 
helpful in clearing up the actual preference of 
choosing the form $q=\exp({\rm i}\theta)$
of deformation parameter. 
It is just this alternative for the choise of 
the deformation parameter $q$ that implies very 
attractive physical interpretation of the $q$-parameter 
as the one that is directly linked to the mixing issue 
of elementary particles, either of bosons                  \cite{Isaev,JPA}
or fermions                                               \cite{NATO,NP_B}.

The paper is organized as follows.
Section 1 contains a sketch of necessary preliminaries 
concerning the two most popular types of $q$-deformed 
oscillators, as well as their two-parameter or 
$qp$-generalization. 
In section 2 we discuss basic points of the approach based 
on the $q$-Bose gas model, along with consideration of 
single particle $q$-distributions.
The remaining two sections are devoted to the properties of
two-particle and three-particle correlation functions, 
and to the main topic of present paper -- the results on 
the multi-particle ($n$-th order) correlations, for the 
algebras of both the $q$-deformed and the $qp$-deformed 
versions of generalized oscillators. 
Details of derivation of basic formulas are relegated 
to the Appendix.


\section{$q$-Deformed and $qp$-deformed oscillators}


We begin with a necessary setup concerning two types
of $q$-deformed oscillators, and also their
two-parameter generalization.



\underline{  $q$-Oscillators of AC type}. \
The $q$-oscillators of the Arik-Coon (or AC-) type
are defined by the relations                               \cite{AC,AFZMP}
\[          
a_i a_j^\dg-q^{\d_{ij}} a_j^\dg a_i=\d_{ij}     \qquad
[a_i,a_j]=[a^\dg_i,a^\dg_j]=0\ 
\]
\beq
[{\cal N}_i,a_j]=-\d_{ij} a_j          \qquad
[{\cal N}_i,a^\dg_j]=\d_{ij} a^\dg_j     \qquad
[{\cal N}_i, {\cal N}_j]=0\ .  
\eeq
were $ -1\le q\le 1 $.
Note that this is the {\it system of independent} $q$-oscillators
as clearly seen at $i\ne j$. 

{}From the vacuum state given by $a_i|0,0,\ldots\ra=0 \ $ for all $i$,
the basis state vectors
\beq   \fl
|n_1,\ldots,n_i,\ldots\ra\equiv
\frac{1}
{\sqrt{\lfloor n_1\rfloor!\lfloor n_2\rfloor!
\cdots\lfloor n_i\rfloor!\cdots}}
(a^\dg_1)^{n_1}(a^\dg_2)^{n_2}\cdots(a^\dg_i)^{n_i}\cdots|0,0,\ldots\ra
\eeq
are constructed as usual, so that
\[                 \fl
a^\dg_i|\ldots,n_i,\ldots\ra =
\sqrt{\lfloor n_i\!+\!1\rfloor }|\ldots,n_i\!+\!1,\ldots\ra      
           \qquad    
a_i|\ldots,n_i,\ldots\ra =
\sqrt{\lfloor n_i\rfloor }|\ldots,n_i\!-\!1,\ldots\ra 
\]
\beq    
{\cal N}_i|n_1,\ldots,n_i,\ldots\ra=n_i|n_1,\ldots,n_i,\ldots\ra 
\eeq
%
%
Here the notation $\lfloor \ldots \rfloor$ for so-called basic numbers
and the corresponding extension of factorial, namely
\beq        \fl                            \qquad 
\lfloor r\rfloor \equiv \frac{1-q^r}{1-q}     \qquad
\lfloor r\rfloor! \equiv
\lfloor 1\rfloor \lfloor 2\rfloor \cdots
\lfloor r-1\rfloor \lfloor r\rfloor           \qquad
\lfloor 0\rfloor!=\lfloor 1\rfloor!=1 
\eeq
are used.
The $q$-bracket $\lfloor A\rfloor$ for an operator $A$
is understood as formal series.
At $q\to 1$, from $\lfloor r\rfloor$ and $\lfloor A\rfloor$
one recovers $r$ and $A$, thus going back to the formulas for the
standard bosonic oscillator.
For the {\it deformation parameter} $q$ such that $-1\le q\le 1$,
the operators $a^\dg_i\ ,$ $a_i$ are conjugates of each other.

For $q\ne 1$, the bilinear $a^\dg_i a_i$ depends  nonlinearly on
the number operator ${\cal N}_i$ :
\beq
a^\dg_i a_i=\lfloor{\cal N}_i\rfloor 
\eeq
so that at $q=1$ the familiar equality $a^\dg_i a_i={\cal N}_i $
is recovered.

\underline{ $q$-Oscillators of BM type}. 
The $q$-oscillators of Biedenharn-Macfarlane (BM)
type are defined by the relations                          \cite{BM,AFZMP}:
\[    \fl      \quad
[b_i,b_j]=[b^\dg_i,b^\dg_j]=0     \qquad
[N_i,b_j]=-\d_{ij} b_j            \qquad
[N_i,b^\dg_j]=\d_{ij} b^\dg_j     \qquad
[N_i, N_j]=0 
\]
\beq
b_i b_j^\dg-q^{\d_{ij}} b_j^\dg b_i=\d_{ij}q^{-N_j}   \qquad
b_i b_j^\dg-q^{-\d_{ij}} b_j^\dg b_i=\d_{ij}q^{N_j}\ . 
\eeq
In this case the extended Fock space of basis state vectors
is constructed in the way similar to the above case, with
the only modification that now we use, instead of basic numbers,
the $q$-bracket and $q$-numbers, namely
\beq
b^\dg_i b_i=[N_i]_q     \qquad   
                        \qquad
[r]_q \equiv \frac{q^r-q^{-r}}{q-q^{-1}} \ .
\eeq
Formulas similar to  (2)-(5) are valid for the
operators $b_i, \ b_j^\dg \ $ if, instead of (4), we now  use
the definition (7) for $q$-bracket.
Clearly, the equality $b^\dg_i b_i = N_i$ holds only in the
``no-deformation'' limit of $q=1$.
For consistency of the conjugation, we put
\beq
q=\exp (i \t)\   \qquad       0 \le \t < \pi \ .
\label{19}
\eeq

    \vspace{1mm}
\underline{$qp$-Oscillators}. 
Besides the $q$-bosons of AC-type and BM-type,
in what follows we will also consider the two-parameter
(or $qp$-) generalization of deformed oscillators
given by the relations                                        \cite{Chakra}
\[
[N^{(qp)},A]=-A   \     \qquad\qquad
[N^{(qp)},A^\dg]= A^\dg\ 
\]
\begin{equation}
A A^\dg - q A^\dg A=p^{N}    \qquad\qquad
 A A^\dg - p A^\dg A=q^{N}
\end{equation}
from which
\begin{equation}
A^\dg A=[\![N^{(qp)}]\!]_{qp}
                   \qquad  {\rm with}   \qquad
[\![X]\!]_{qp} \equiv \frac{ q^{X}-p^{X} }{ q-p } \ .
\end{equation}
In this definition we have shown only one mode, although similarly
to (1) and (6), in what follows we will deal with the system of
independent (that is, mutually commuting) copies/modes of
the $qp$-deformed oscillator. Note that $X$ in (10) can be 
either a number or an operator.
Clearly, putting $p=1$ immediately leads us to the AC-case
while putting $p=q^{-1}$ reduces to the BM-type of $q$-bosons.


\section{ Statistical $q$-distributions }


For the dynamical multiparticle (say, multi-pion or multi-kaon)
system, we consider the model of ideal gas of $q$-bosons 
(IQBG) by taking the free, or non-interacting, 
Hamiltonian in the form                               \cite{Vokos,AG,Man'ko}
\beq
H=\sum_i{\omega_i {\cal N}_i}
\eeq
where $\omega_i=\sqrt{m^2+{\bf k}_i^2}\ $, ${\cal N}_i$ is
the number operator given in (5) or (7) or (10), and the
subscript $'i'$ labels different modes.
Let us note that among a variety of
possible choices of Hamiltonians, the choice (11)
is the unique truly non-interacting one, which
possesses an additive spectrum, see                        \cite{AG,Vokos}.
Clearly, it is assumed that $3$-momenta of
particles take their values from a discrete set (i.e. the system
is contained in a large finite box of volume $\sim L^3$).

To obtain basic statistical properties, one evaluates
thermal averages
\[
\la A \ra=\frac{{\rm Sp}(A\rho)}{{\rm Sp}(\rho)}   \qquad
\rho=e^{-\b H} 
\]
where $\b=1/T$ and the Boltzmann constant is set equal to 1.
Calculating, say, in the case of AC-type $q$-bosons
the thermal average $\la q^{{\cal N}_i} \ra $,
with ${\cal N}_i$ from (5), with respect to the 
chosen Hamiltonian (11) we obtain
\beq
\la q^{{\cal N}_i} \ra =
\frac{e^{\b\omega_i}-1}{e^{\b\omega_i}-q}
\eeq
and the distribution function
(recall that $ -1\le q\le 1 $) is found as                \cite{AG,Vokos}:
\beq
\la a_i^\dg a_i \ra=\frac{1}{e^{\b\omega_i}-q}\ .
\eeq
Usual Bose-Einstein distribution corresponds to
the no-deformation limit of $q\to 1$.
In the particular cases $q=-1$ or $q=0$ the distribution
function (13) yields respectively Fermi-Dirac or
classical Boltzmann ones.
Note that this coincidence is rather a formal one: the defining
relations (1) at $q=-1$ or $q=0$ differ from those for the system
of fermions or the non-quantal (classical) system.
The formal coincidence of equation (13) at $q=-1$ with the
Fermi-Dirac distribution can be interpreted                     
in terms of impenetrability (or hard-core) property
of such bosons.
The difference with the system of genuine fermions lies
in {\it commuting} (versus truly fermionic anticommuting) of
{\it non-coinciding} modes at $q=-1$, see (1).

   Now consider BM-type of $q$-bosons.
The Hamiltonian is chosen again as that of IQBG, but now 
with the number operator given in (7), i.e.,
\beq
H=\sum_i{\omega_i  N_i}\ .
\eeq
Calculation of $\la q^{\pm N_i} \ra $ yields
$ \la q^{\pm N_i} \ra =
{(e^{\b\omega_i}-1)}{(e^{\b\omega_i}-q^{\pm 1})^{-1}} . $
Then, from the formula
$\la b_i^\dg b_i \ra = 
{(e^{\b\omega_i}-q)^{-1}}\la q^{-N_i} \ra $
the expression for the $q$-deformed distribution 
function (note that $q+q^{-1}=[2]_q=2\cos\t$) does 
follow (see also                                        \cite{AG,Vokos}):
\beq
\la b_i^\dg b_i \ra=\frac{e^{\b\omega_i}-1}
{e^{2\b\omega_i}-2\cos\theta\ e^{\b\omega_i}+1} \ .
\eeq
Although the deformation parameter $q$ is taken as
{\it complex} one according to (8), the explicit expression
(15) for the $q$-distribution function shows that it is real.

It is easily seen that the shape of the function
$f({\bf k})\equiv \la b^\dg b \ra ({\bf k}) $ in (15) is such that
the $q$-deformed distribution function with $q\ne 1$ is
intermediate relative to the other two curves, the standard
Bose-Einstein distribution function and the classical Boltzmann one
(the same is also evident for the above $q$-distribution function (13)
 of the AC-type $q$-bosons).
That is, the deviation of the $q$-distribution (15) from
the quantum Bose-Einstein distribution goes, when $q$ goes away
from the no-deformation limit $q=1$, in the ``right direction'',
towards the classical Boltzmann one.


\section{ Two- and three-particle correlations of $q$-bosons }
\label{sec4}


Although the formulas for two-particle correlation
functions have been obtained earlier                         \cite{AGI-1},
we recall them here for the sake of
more complete exposition.
In the remaining part of this section some new 
results will be presented.
So, we consider two-particle correlations first
with the AC- type of $q$-bosons.
Starting with the identity
\[       \fl
a_i^\dg a_j^\dg a_k a_l \!-\! q^{-\d_{ik}-\d_{il}}  a_j^\dg a_k a_l a_i^\dg
=[a_i^\dg , a_j^\dg] a_k a_l + a_j^\dg [a_i^\dg , a_k]_{q^{-\d_{ik}}} a_l
+q^{-\d_{ik}}  a_j^\dg a_k [a_i^\dg,a_l]_{q^{-\d_{il}}} 
\]
where $[X,Y]_\kappa\equiv XY\!-\!\kappa YX$,
by taking thermal averages of its both sides we find
\[
\la a_i^\dg a_j^\dg a_k a_l\ra=\frac{e^{\b\omega_i}-q}
{q^{1-\d_{ik}-\d_{il}}e^{\b\omega_i}-q}(
\la a_j^\dg a_l \ra\la a_i^\dg a_k\ra+q^{-\d_{ij}}
\la a_j^\dg a_k \ra\la a_i^\dg a_l \ra) \ .
\]
For coinciding modes this leads to the formula
\beq
\la a_i^\dg a_i^\dg a_i a_i\ra=\frac{1+q}
{(e^{\b\omega_i}-q)(e^{\b\omega_i}-q^2)}\ .\ \
\eeq
{}From the last relation and the $q$-distribution (13)
the ratio under question (called intercept) does result:
\beq
\lambda_i \equiv \frac{\la a_i^\dg a_i^\dg a_i a_i\ra}
{\la a_i^\dg a_i \ra^2} - 1 =
- 1 + \frac{(1+q)(e^{\b\omega_i}-q)}
{e^{\b\omega_i}-q^2} =
q\frac{e^{\b\omega_i}-1}{e^{\b\omega_i}-q^2}\ .
\eeq
%
Note that in the non-deformed limit $q\to 1$
the value $\lambda_{\rm BE}=1$, proper for Bose-Einstein statistics,
is correctly reproduced from equation (17).
This obviously corresponds to the Bose-Einstein distribution
contained in (13) at $q\to 1$.
The quantity (intercept) $\lambda$ is important since it can be directly
confronted with empirical data.
In this respect, let us note that there exists  a direct asymptotic
relation $\lambda=q$, which corresponds to the limit of large momentum
or low temperature (in that case $\b\omega\to\infty$).


Now we go over to the Biedenharn-Macfarlane
$q$-oscillators (6) and find the formula for the
monomode two-particle correlations, i.e. for identical
particles with coinciding momenta.
From the relation
\[
\la b_i^\dg b_i^\dg b_i b_i\ra -q^{2} \la b_i^\dg b_i b_i b_i^\dg\ra
=-\la b_i^\dg b_i q^{N_i} \ra (1+q^2) \,
\]
valid for the monomode case at hand, we deduce
\[
\la b_i^\dg b_i^\dg b_i b_i \ra=\frac{1+q^2}{q^2 e^{\b\omega_i}-1}
\la b_i^\dg b_i q^{N_i}\ra \ .
\]
Evaluation of the thermal average in the r.h.s. yields
$\la b_i^\dg b_i q^{N_i}\ra = q /(e^{\b\omega_i}-q^2).$
Using this we find the expression for two-particle
distribution, namely
\beq
\la b_i^\dg b_i^\dg b_i b_i\ra=\frac{2\cos\theta}
{e^{2\b\omega_i}-2\cos(2\theta)e^{\b\omega_i}+1}\ .
\eeq
Then, the desired formula for the intercept of two-particle
correlations of the BM-type $q$-bosons, with the notation 
$t_i\equiv\cosh(\b\omega_i)-1$, reads 
\beq
\lambda_i = - 1 + \frac{\la b_i^\dg b_i^\dg b_i b_i\ra}
{(\la b_i^\dg b_i \ra)^2}=
\frac{2\cos\theta(t_i+1-\cos\theta)^2}
     {t_i^2+2(1-\cos^2\theta)t_i} 
\eeq
and again is a real function.

\begin{center}
  { {\em Three-particle correlations of the} $q$-{\em bosons
                             of AC-type} }
\end{center}

Derivation of three-particle correlation functions
proceeds analogously to the 2-particle case.
Considering the $q$-deformed oscillators of AC-type        
we start with the easily verifiable identity
\[      
a_j^\dg a_k^\dg a_l a_m a_s a_i^\dg =
a_j^\dg a_k^\dg a_l a_m [a_s, a_i^\dg]_{q^{\delta_{is}}} +
\]
\[                    \fl     \ \ \
 + q^{\d_{is}}
\left\{
a_j^\dg a_k^\dg a_l ( [a_m,a_i^\dg]_{q^{\delta_{im}}} ) a_s
+ q^{\d_{im}}
\left (
a_j^\dg a_k^\dg ( [a_l,a_i^\dg]_{q^{\delta_{il}}} ) a_m a_s
+ q^{\d_{il}} a_j^\dg a_k^\dg a_i^\dg a_l a_m a_s
\right )
\right\}
\]
and take thermal averages of both its sides.
  This leads to the equality 
\[
\la   a_i^\dg a_j^\dg a_k^\dg a_l a_m a_s  \ra   
= \frac{e^{\b\omega_i}-q}{e^{\b\omega_i}-q^{\d_{is}+\d_{im}+\d_{il}}}
\left(
\la a_j^\dg a_k^\dg a_l a_m\ra\la a_i^\dg a_s\ra       \right.
\]
\[     \left.    
+ q^{\d_{is}}
\la a_j^\dg a_k^\dg a_l a_s\ra\la a_i^\dg a_m
\ra
+  q^{\d_{is}+\d_{im}}
\la a_j^\dg a_k^\dg a_m a_s\ra\la a_i^\dg a_l
\ra    \right)
\]
which in view of
$\la a_i^\dg a_j\ra=\delta_{ij}\la a_i^\dg a_i\ra
=\delta_{ij}/(e^{\b\omega_i}-q)$, cf. (13),
in the monomode $i=j=k=l=m=s$ case yields:
\beq
\la a_i^\dg a_i^\dg a_i^\dg a_i a_i a_i\ra =
\frac{(1+q)(1+q+q^2)}
{(e^{\b\omega_i}-q)(e^{\b\omega_i}-q^2)(e^{\b\omega_i}-q^3) }\ .\ \
\eeq
{}From the latter relation, dividing it by $\la a_i^\dg a_i\ra^3$,
we derive the desired expression for the intercept
(or strength) $\lambda^{(3)}$ of 3-particle correlation 
function (we drop the '$i$'):  
\beq
\lambda_{\rm {AC}}^{(3)}
\equiv \frac{ \la a^{\dg  3} a^3\ra }
            { \la a^\dg a\ra^3 } - 1 =
\frac{(1+q)(1+q+q^2)(e^{\b\omega}-q)^2}
{(e^{\b\omega}-q^2)(e^{\b\omega}-q^3)} - 1 .
\eeq

In a similar manner, it is possible to derive the
(intercept of) 3-particle correlation function for the system
of BM-type $q$-bosons.
However, intead of doing this, in the next section we will
derive the most general results for both 3- and $n$-particle,
$\ n>3$, correlation functions in the two-parameter
(i.e. $qp$-deformed) extension of bosons, from which
the desired formulae for the BM-type of
$q$-bosons will follow as particular cases.


\section{ $n$-Particle correlations:  $q$-bosons and $qp$-bosons }
\label{sec5}


As extension of equations (16), (20), it is not difficult 
to derive, using method of induction, the following general 
result for the $n$-particle monomode distribution functions of 
AC-type $q$-Bose gas:
\beq           \fl
\la ( a_i^\dg )^n  ( a_i )^n \ra =
\frac{ \lfloor n\rfloor! }
{\prod_{r=1}^n (e^{\b\omega_i}-q^r) }        \qquad
\lfloor m\rfloor \equiv 
\frac{1-q^m}{1-q}=1+q+q^2+...+q^{m-1} \ .
\eeq
From this expression the desired formula for the intercepts
$\lambda^{(n)} \equiv
\frac{ \la a^{\dg  n}~ a^n\ra } { \la a^\dg a\ra^n } - 1 $
of $n$-particle correlations of AC-type $q$-bosons
immediately follows (with '$i$' dropped):
\beq
\lambda^{(n)}_{\rm {AC}}= - 1 +
\frac{ \lfloor n\rfloor! \ (e^{\b\omega}-q)^{n-1} }
{ \prod^n_{r=2}(e^{\b\omega}-q^r) }.
\eeq
In the asymptotics of $\b\omega\to\infty$
(i.e., for very large momenta or, at fixed momentum,
for very low temperature) the result depends only
on the deformation parameter:
\[             
\lambda^{(n) \ {\rm asympt}}_{\rm {AC}} = -1 +
\lfloor n\rfloor! = -1+ \prod^n_{k=1}\biggl(\sum^k_{r=0}q^r\biggr)
\]
\beq
= (1+q)(1+q+q^2)\cdots(1+q+\ldots+q^{n-1}) -1 .
\eeq
This remarkable fact can serve as the test one when
confronting the developed approach with the numerical
data for pions and kaons extracted from the experiments
on relativistic heavy ion collisions.

Now we come to the base point.

\begin{center}
{\em  Extension to $qp$-bosons.}
\end{center}

The above results admit direct extension to the case of
the two-parameter deformed (or $qp$-)oscillators and thus
to the $qp$-Bose gas model.
For this, we use in analogy with (11) and (14)
the Hamiltonian
\beq
H=\sum_i{\omega_i ~ N^{(qp)}_i}\ .
\eeq
With (25), the expression for general $n$-particle
distribution functions is obtained 
(see Appendix for its derivation) as
\beq
\la ( A_i^\dg )^n  ( A_i )^n \ra =
\frac{   [\![n]\!]_{qp}! \ (e^{\b\omega_i}-1) }
{\prod_{r=o}^n ( e^{\b\omega_i} - q^r p^{n-r} ) } 
\eeq    
\[
[\![m]\!]_{qp}\equiv\frac{q^m-p^m}{q-p}      \qquad
[\![m]\!]_{qp}!=[\![1]\!]_{qp}[\![2]\!]_{qp} \cdots
[\![m-1]\!]_{qp}[\![m]\!]_{qp} \ .
\]
In the particular cases $n=1$ and $n=2$ (note that
$[\![2]\!]_{qp}=p+q$) this obviously yields the formulas
\[
\la  A_i^\dg A_i  \ra =
    \frac{ (e^{\b\omega_i}-1) }
       { (e^{\b\omega_i}-p) (e^{\b\omega_i}-q) } 
\]
\[
\la ( A_i^\dg )^2  ( A_i )^2 \ra =
\frac{ (p+q) \ (e^{\b\omega_i}-1) }
{(e^{\b\omega_i}-q^2)(e^{\b\omega_i}-pq)(e^{\b\omega_i}-p^2)}
\]
(remark that the latter two formulas were also found in        \cite{Daoud}).

From (26), after dividing it by $\la  A_i^\dg A_i \ra^n$, 
the most general result for the $n$-th order $qp$-deformed
extension of the intercept $\lambda^{(n)}$, omitting the '$i$', 
follows as
\begin{equation}     \fl        \ \ \ \ 
\lambda^{(n)}_{q,p}  \equiv
 \frac{{\langle A^{\dag n} A^{n} \rangle}}{{\langle
A^{\dag} A \rangle}^{n}} - 1 =  
[\![n]\!]_{qp}! \ 
\frac{(e^{\b\omega}-p)^n (e^{\b\omega}-q)^n}
{(e^{\b\omega}-1)^{n-1}
\prod_{k=0}^n(e^{\b\omega}-q^{n-k}p^k)} - 1 
\end{equation}
which constitutes our main result.
This provides generalization not only to the case of
$n$-th order correlations but also to the two-parameter
($qp$-)deformation.

Let us give the asymptotical form of intercepts in this
most general case, $\lambda^{(n)}_{q,p}$:
\begin{equation}
\lambda^{(n), \ {\rm asympt} }_{q,p}
= - 1 + [\![n]\!]_{qp}!
= - 1 + \prod^n_{k=1}\biggl(\sum^k_{r=0}q^r p^{k-r}\biggl) \ .
\end{equation}
As we see, for each case of deformed bosons (the
AC-type, the BM-type, and their $qp$-generalization)
the asymptotics of the $n$-th order intercept takes the form
of the corresponding generalization of the usual $n$-factorial
(the latter yields pure Bose-Einstein $n$-particle
correlation intercept).

Finally, let us specialize the obtained formulae
to the case of $q$-bosons of BM type for $n=3$, that is
\begin{equation}\lambda^{(3)}_{\rm {BM}} = - 1 +
\frac{[2]_q[3]_q\ (e^{2\b\omega}-2e^{\b\omega}\cos\!\theta +1)^2}
     {(e^{\b\omega}-1)^2 (e^{2\b\omega}-2e^{\b\omega}\cos(3\theta)+1)}\ 
\end{equation}
\begin{equation}
\lambda^{(3), \ {\rm asympt} }_{\rm {BM}} = 
-1+[2]_q[3]_q =-1+ 2\cos\theta\ (2\cos\theta -1)(2\cos\theta +1).
\end{equation}
In conclusion we note that it would be of great interest and
importance to make a detailed comparative analysis of
the obtained results with the existing data for 3-particle
correlations of pions and kaons produced and registered in
the experiments on relativistic heavy ion collisions,
with the goal to draw some implications concerning immediate
physical meaning and admissible values of the deformation
parameters $p,q$. Details of such analysis will be presented
elsewhere.

     \appendix
     \section*{Appendix}
 \setcounter{section}{1}


Here we derive the general formula, see (26), for the (monomode)
$n$-particle $pq$-bosonic distribution functions:
\begin{equation}
\langle a^{\dag n} a^{n} \rangle=\frac{[\![n]\!]_{qp}!
(e^{\b\omega}-1)}{\prod_{r=0}^{n} (e^{\b\omega}-p^r q^{n-r})} \ .
\end{equation}
For convenience, in (A.1) and below, we drop
 the mode-labelling subscript "$i$", and  use  
$a^\dag, a, N $ instead of  $A^\dag,\ A,\ N^{(qp)}$
respectively.   The proof proceeds in few steps.
First let us derive the recursion relation
\begin{equation}
\langle a^{\dag n} a^{n} \rangle  =
 \langle a^{\dag n-1} a^{n-1}p^{N} \rangle
 \frac{ [\![n]\!]_{qp} } { (e^{\b\omega} -q^n) p^{n-1} }\ .
\end{equation}
For this, we use $pq$-deformed commutation relations
and evaluate the thermal averages:
\[            
\langle a^{\dag n} a^{n} \rangle  =
{ \langle a^{\dag n-1} a a^{\dag} a^{n-1} \rangle } \frac{1}{q} -
{ \langle a^{\dag n-1} p^{N}
a^{n-1} \rangle }  \frac{1}{q}  
\]
\[
= { \langle a^{\dag n-1} a a^{\dag} a^{n-1} \rangle }  \frac{1}{q} -
{ \langle a^{\dag n-1} a^{n-1} p^{N} \rangle } \frac{1}{q p^{n-1}}
\]
\[
= { \langle a^{\dag n-1} a^2 a^{\dag} a^{n-2} \rangle }  \frac{1}{q^2}
   - \frac{1}{q}
     \biggl( \frac{1}{p^{n-1}}+ \frac{1}{q p^{n-2}}  \biggr)
{ \langle a^{\dag n-1} a^{n-1} p^{N} \rangle }  = \ldots
\]
\[
=
 { \langle a^{\dag n-1} a^{n}  a^{\dag} \rangle  } \frac{1}{q^{n}}  -
  \frac{1}{q} \biggl ( \frac{1}{p^{n-1}} + \frac{1}{q p^{n-2}} +...+
\frac{1}{q^{n-1}} \biggr )
 { \langle   a^{\dag n-1} a^{n-1} p^{N}
 \rangle }
\]
\[
=
  {\langle a^{\dag n-1} a^{n} a^{\dag} \rangle
} \frac{1}{q^{n}}  - { \langle a^{\dag n-1} a^{n-1} p^{N}
\rangle } \frac{[\![n]\!]_{qp}}{q^n p^{n-1}}
\]
\begin{equation}                 
=    {\langle a^{\dag n} a^{n} \rangle
} \frac{e^{\b\omega}}{q^n} - {\langle a^{\dag n-1} a^{n-1} p^{N}
\rangle } \frac{[\![n]\!]_{qp}}{q^n p^{n-1}} \ .
\end{equation}
From this the equation (A.2) readily follows.
After $k$-th iteration of this
procedure we find

\begin{equation}                   \fl
{ \langle a^{\dag n-k} a^{n-k} p^{kN} \rangle } =
{ \langle a^{\dag n-(k+1)} a^{n-(k+1)} p^{(k+1)N} \rangle }
{ \frac{[\![n-k]\!]_{qp}}{(e^{\b\omega}-q^{n-k}p^k)p^{n-(2k+1)}} } \ . \
\end{equation}
Indeed,
\[                              \fl
\langle a^{\dag n-k} a^{n-k} p^{kN} \rangle =
\langle a^{\dag n-(k+1)} a a^{\dag} a^{n-k-1} p^{kN} \rangle  \frac{1}{q}
  -  \langle a^{\dag n-(k+1)} p^N a^{n-(k+1)} p^{kN} \rangle  \frac{1}{q}
 = 
{...}
\]
\[                              
=  \langle a^{\dag n-k} a^{n-k} p^{kN} \rangle
\frac{e^{\b\omega}}{q^{n-k} p^k}   
- \frac{1}{q} \biggl( \frac{1}{p^{n-1-k}} + \frac{1}{q
p^{n-2-k}} +...
\]
\[
+ \frac{1}{q^{n-k-1}} \biggr){ \langle a^{\dag n-(k+1)}
a^{n-(k+1)} p^{(k+1)N} \rangle }
\]
\begin{equation}                \fl
=
 \langle a^{\dag n-k} a^{n-k} p^{kN} \rangle
\frac{ e^{\b\omega} }{ q^{n-k} p^k }
  -  \langle a^{\dag n-(k+1)} a^{n-(k+1)} p^{(k+1)N} \rangle
 \frac{ [\![n-k]\!]_{qp} }{  q^{n-k} p^{n-(2k+1)} }
\end{equation}
that is equivalent to the formula (A.4).
Applying this formula step by step $n$ times
yields the relation
\begin{equation}
\langle a^{\dag n} a^{n} \rangle 
= \frac{[\![n]\!]_{qp}!}{\prod_{r=0}^{n-1} (e^{\b\omega}-p^r q^{n-r})
\prod_{k=0}^{n-1}p^{n-(2k+1)}} {\langle p^{nN} \rangle } \ .
\end{equation}
From the latter, with the account of
\begin{equation}
{\langle p^{nN}\rangle}
=\frac{e^{\b\omega}-1}{e^{\b\omega}-p^n} \qquad\qquad
%
\prod_{k=0}^{n-1}p^{n-(2k+1)}=1 
\end{equation}
we finally arrive at the desired formula (A.1) for
the higher order ($n$-particle) monomode distribution
functions of the model of $qp$-Bose gas.



\Bibliography{26}

\bibitem{Bied}
Biedenharn L 1990 in {\em Group Theoretical Methods in Physics} 
(Moscow:  V V Dodonov and V I Man'ko eds), Lecture Notes
in Physics {\bf 382} ( Berlin: Springer) p 147

\bibitem{Zachos}   Zachos C    1990
{\em Proc. Argonne Workshop on Quantum Groups}
(Singapore: World Scientific T Curtright, D Fairlie, and C Zachos eds.) 

\bibitem{Chang}     Chang Z    1995
{\em Phys. Rep.} {\bf 262} 137

\bibitem{Chai}
Chaichian M, Gomez J F and Kulish P  1993
{\em Phys. Lett.} {\bf 311B} 93

\bibitem{Fair} Fairlie D and Nuits J 1995
{\em Nucl Phys B} {\bf 433} 26 

\bibitem{JPA}  Gavrilik A M   1994
{\em J. Phys. A} {\bf 27} L91

\bibitem{GI}  Gavrilik A M and Iorgov N Z 1998
{\em Ukrain. J. Phys.} {\bf 43} 1526    \\
 ( Gavrilik A M and Iorgov N Z 1998 
            {\it Preprint} {\tt hep-ph/9807559})

\bibitem{NP_B}   Gavrilik A M  2001
 {\em Nucl. Phys. B (Proc. Suppl.)}  {\bf 102/103} 298  \\
  ( Gavrilik A M  2001 {\it Preprint} {\tt hep-ph/0103325})

\bibitem{AGI-1}
 Anchishkin D V, Gavrilik A M and Iorgov N Z 2000
{\em Europ. Phys. Journ.  A} {\bf 7} 229  \\
( Anchishkin D V, Gavrilik A M and Iorgov N Z 2000
{\it Preprint}  nucl-th/9906034)

\bibitem{Heinz99}
Heinz U and Jacak B V   1999
{\em Annu. Rev. Nucl. Part. Sci.} {\bf 49} 529

\bibitem{AGI-2} 
  Anchishkin D V, Gavrilik A M and Iorgov N Z      2000
  {\em Mod.~Phys.~Lett. A} 1{\bf 5} 1637   \\
 ( Anchishkin D V, Gavrilik A M and Iorgov N Z 2000
   {\it Preprint}  {\tt hep-ph/0010019})

\bibitem{AGP}
Anchishkin D V, Gavrilik A M  and Panitkin S    2001
Transverse momentum dependence of intercept parameter $\lambda$
of two-pion (-kaon) correlation functions in $q$-Bose gas model
     {\it Preprint} {\tt hep-ph/0112262}.

\bibitem{Adams}        
  Adams J et al.      2003
Three-pion HBT correlations in relativistic heavy ion collisions
from the STAR experiment  
      {\it Preprint}   {\tt nucl-ex/0306028}.
 
\bibitem{Heinz97}  
 Heinz U and Zhang Q H     1997 
{\em Phys. Rev. C} {\bf 56} 426

\bibitem{Csorgo}    
 Csorg\"o  T         2002
{\em Heavy Ion Physics} {\bf 15} 1

\bibitem{Isaev}    
 Isaev A P and Popowicz Z    1992
{\em Phys. Lett. B} {\bf 281} 271

\bibitem{NATO}    
Gavrilik A M   2001 
in {\em Proc. of NATO Advanced Studies Workshop "Noncommutative 
Structures in Mathematics and Physics"}, 
Kluwer, Dordrecht, p. 344   \\
(Gavrilik A M 1998 {\it Preprint} {\tt hep-ph/0011057})

\bibitem{AC}
Coon D D, Yu S and Baker M   1972
    {\em Phys.  Rev. D} {\bf 5} 1429     \\
Arik M and Coon D D   1976
   {\em J. Math. Phys.}  {\bf 17} 524

\bibitem{AFZMP}
M.~Arik, Z.\ Phys.\ {\bf C\ 51}, 627 (1991).
D.~Fairlie and C.~Zachos, Phys.\ Lett.\ {\bf 256B}, 43 (1991).
S.~Meljanac and A.~Perica,
Mod.\ Phys.\ Lett.\ {\bf A 9}, 3293 (1994).

\bibitem{BM}
Macfarlane A J 1989  {\em  J. Phys. } {\bf A 22} 4581   \\
Biedenharn L C 1989 {\em J.  Phys. } {\bf A 22} L873
 
\bibitem{Chakra}
Chakrabarti R and Jagannathan R     1991
{\em J.~Phys. A} {\bf 24} L711

\bibitem{AG}
Altherr T and Grandou T     1993
{\em Nucl. Phys.} {\bf B402} 195   

\bibitem{Vokos}
Vokos S and Zachos C     1994
{\em Mod. Phys. Lett.} {\bf A 9} 1

\bibitem{Man'ko}
Man'ko V I, Marmo G, Solimeno S and Zaccaria F     1993
{\em Phys. Lett.} {\bf 176 A} 173     

\bibitem{Daoud} 
Daoud M and Kibler M              1995
{\em Phys. Lett. A}  {\bf 206} 13
 
\endbib
\end{document}